# Comment on "Comparative study of beta-decay data for eight nuclides measured at the Physikalisch-Technische Bundesanstalt" [Astropart. Phys., 50, 47-58]

Ole Nähle*, Karsten Kossert

*Physikalisch-Technische Bundesanstalt (PTB), Bundesallee 100, 38116 Braunschweig, Germany*

**Abstract**

*We would like to comment on a recent paper by Sturrock et al. [1] in which the authors analyze decay data acquired by an ionization chamber in our institute. They interpret the variations in the data as solar-driven changes in the decay rates of the radionuclides under study. In brief we would like to discuss and elucidate the properties and the origin of the data used by the authors and explain why these data are not a sound basis for claiming evidence for new physics.*

**Introduction**

In a recent paper, Sturrock et al. [1] make use of experimental data measured at PTB claiming evidence for new physics. The data were acquired with an ionization chamber which is a secondary standard for the unit becquerel in Germany. The set-up is not a dedicated high-precision experiment to detect changes in decay rates of radionuclides on a per mille level. We would like to highlight the origin of our data and explain why the interpretations published in [1] are false.

**Ionization chamber measurements at PTB**

Ionization chambers are used at PTB to calibrate radioactive sources in terms of activity concentration. It has been known for decades that gas detectors and the corresponding readout electronics are influenced by environmental parameters like temperature, pressure and humidity. These parameters, by the way, are strongly correlated to and to a great extent caused by the rotation of the Earth around the Sun. Therefore, a correlation (but not causality!) to the solar neutrino flux is to be expected. The known instability of the instruments is also the reason why several laboratories make use of measurements against long-lived reference sources [2]. In our laboratory $^{226}$Ra with a half-life of 1600(7) years is measured and the results are used to reduce the effect of efficiency changes for other isotopes.

Driven by the repetitive use of our data by other groups trying to find solar-induced oscillations, we started to monitor environmental data in our laboratory. Figure 1 shows these parameters and, additionally, the ionization current induced by a $^{226}$Ra source. It can be seen that there is a small oscillation with a frequency of one year in these current measurements and an oscillation with the same frequency but higher amplitude for temperature and relative humidity. When discussing oscillations with an amplitude of less than one per mille, first one has to think about thermal changes of the ionization chamber resulting in geometry and efficiency changes. In addition, the electrical properties of the connecting cables and the Keithley 6517A electrometer used to measure the ionization current are influenced by changes in temperature. The manufacturer of the electrometer claims a temperature coefficient of 0.1 %/°C of the reading in the relevant current ranges [3]. Given the fact that the temperature in our laboratory changes annually between 18 °C and 24 °C, a relative change of the measured current of up to 0.6 % in the worst case could be expected. It seems fallacious to discuss solar neutrino effects while ignoring these obvious causal relationships.

*Corresponding author, e-mail: ole.j.naehle@ptb.de

**Current-variability data**

The variability of the data is on the one hand caused by the process of ionization and charge collection in the ionization chamber. It is obvious that this variability depends on the ionization current which in turn is determined by the activity and the decay properties, e.g. emission probabilities and energies, of the radioactive source under study. The ionization current also depends on chamber properties like gas filling, gas pressure and the design of the charge collecting electrodes. Moreover, the sensitivity of our electrometer to noise depends on the measuring range used. Figures 1d and 1e show the decay-corrected ionization current of the exact same $^{226}$Ra source measured in two separate ionization chambers. This $^{226}$Ra source is routinely measured as a standard and the same electrometer is used to read the current values. The variability of the current readings differs by about a factor of 5 but it would be venturesome to attribute more than detector and detector readout properties to this fact. It should be noted that the ionization currents induced by the eight radioactive sources which are discussed by Sturrock et al. [1], vary by more than one order of magnitude and, thus, differences in variability must be expected.

Another source of variability is the reproducibility of the source position in the chamber, which affects the response of the detector in particular for low energies. Since our system is run by an automated sample changer there are some tolerances in the positioning of the source holder. For low-energy photons and even worse beta particles, there is a strong geometry dependence of the measured current. Moreover, for $^{133}$Ba in particular, we know that our solution is chemically unstable and in this sense "$^{133}$Ba is anomalous". Given the fact that the ionization current is rather small, a higher variability must be expected.

From these arguments one should conclude that there is no scientific value in Figs. 34 to 49 in [1], except for the use as a monitoring tool for the stability of the experimental set-up.

**Analysis of current measurements**

The authors describe a detrending of the data applying a minimization procedure according to Eqs. 2 and 3 in [1]. In fact, this corresponds to a fit of the half-life. Unfortunately the authors do not quote the half-lives obtained, but it must be expected that these differ significantly from the commonly accepted values [4]. Therefore, it must be stressed that Figs. 1 to 8 in [1] are not representative of the quality of our raw data and suggest that our data quality is better than it actually is. Figure 2 shows the same measurements as Fig. 7 in [1] but using the correct half-life of $^{226}$Ra for decay correction. There is a trend in the data of 0.25 % over the period shown caused by detector instabilities. We would like to emphasize that our ionization chambers are not capable of proving effects on a per mille level and that these measurements were never meant to be used for this kind of analysis. It should be noted that the ionization currents must not be equated with decay rates.

Recently, we published the results of our dedicated $^{36}$Cl measurements which give no evidence for a solar-driven variation of the decay rates [5]. Therefore, we conclude that any solar influence on decay rates is very small at best. Thus, sophisticated and dedicated experiments must be made to prove these subtle effects are real. New results from our laboratory based on absolute methods for $^{90}$Sr also disprove any solar influence on decay rates [6].

**Summary**

The work of Sturrock et al. highlights the problems evolving when raw data acquired by one group of experimentalists are analyzed by another group without expert knowledge and experience of the equipment used. The problems grow even worse when the both groups work in different fields, radionuclide metrology and astrophysics in this case. There was no scientific discussion between the authors and PTB in the interpretation of the findings claimed. Therefore we would like to emphasize that we strongly disagree with the conclusions drawn in the paper although the title and the acknowledgements to a former PTB staff member might suggest the opposite.

The findings claimed by Sturrock et al. [1] could easily be explained by detector and readout instabilities [6]. If there were an effect of solar neutrinos on the decay rates, special experiments would need to be undertaken. Using data arbitrarily found in literature without taking the experimental conditions into account, will result in bogus findings and can never be proof of "new physics".

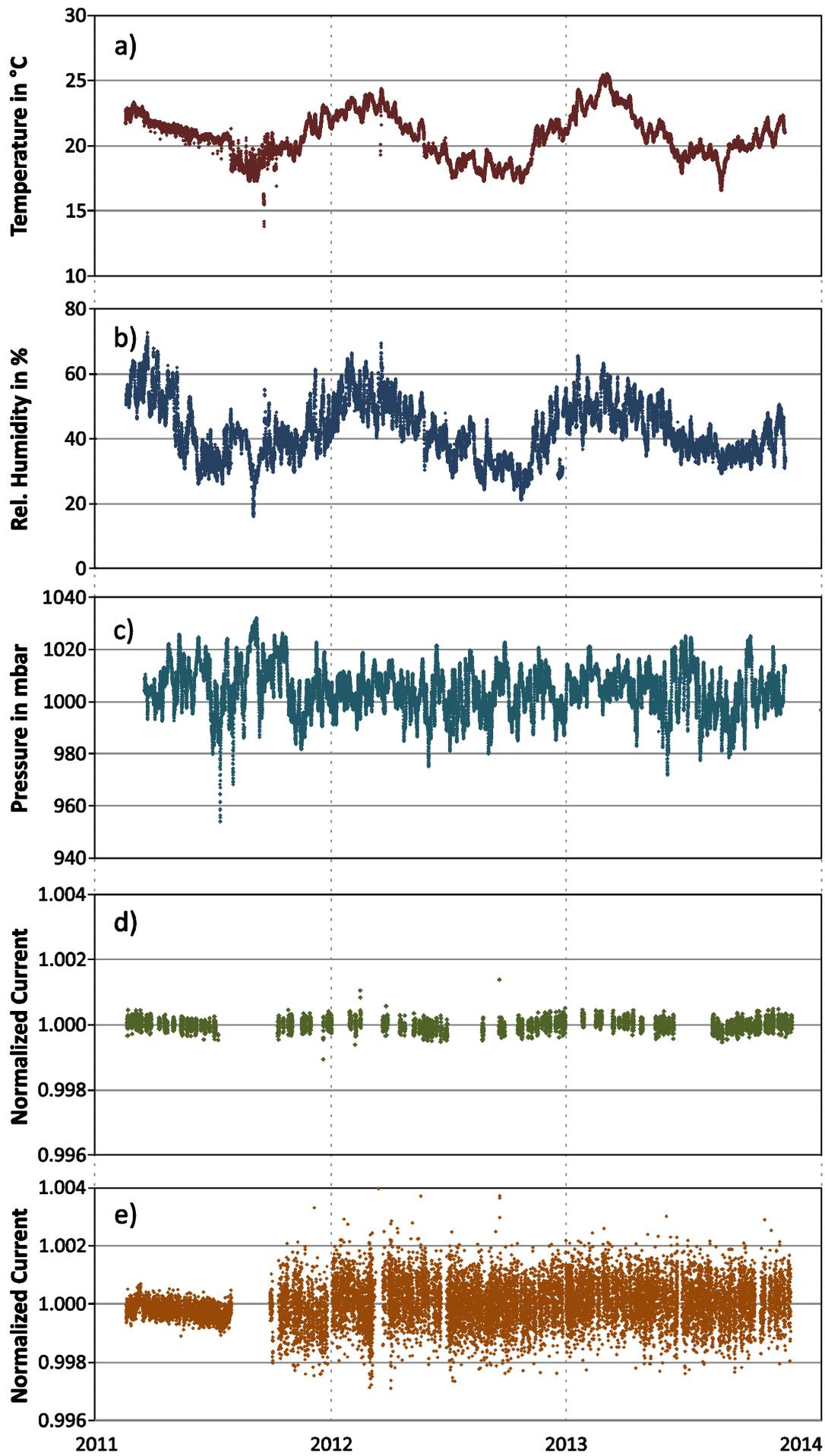

Fig. 1: Experimental conditions after refurbishing the laboratory. Temperature a), relative humidity b) and air pressure c) are monitored every 10 minutes. The lower two plots show the normalized ionization current, averaged over one day, of the same $^{226}$Ra source in two ionization chambers measured by the same electrometer. The raw data were corrected for background and radioactive decay; for one chamber with a minor gas leak, a linear decrease in efficiency was corrected in d). The data in [1] are measured by the other – gas-tight – chamber e).

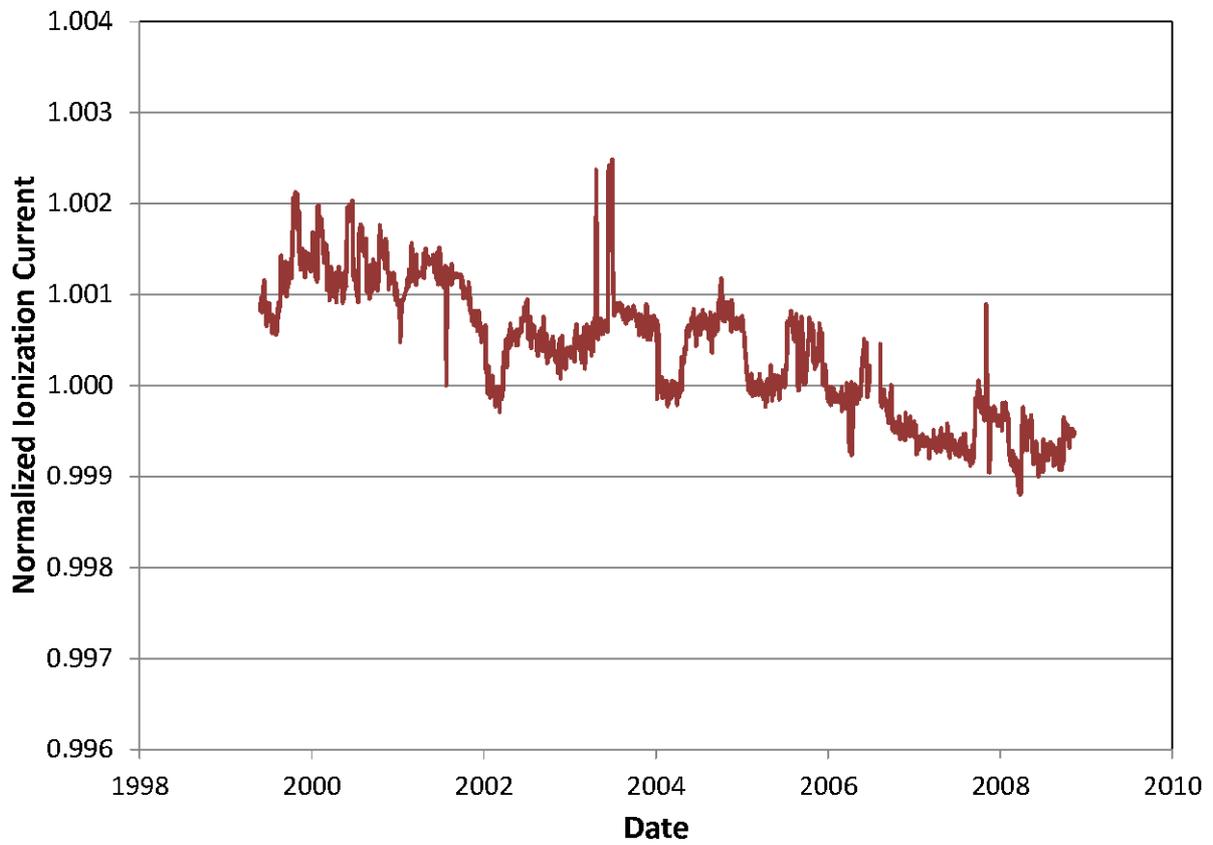

Fig. 2: Experimental data without the arbitrary detrending as shown in Fig. 7 of [1]. The decay correction applied uses a half-life of 1600(7) years as commonly agreed in radionuclide metrology [4].